\documentclass[aps,prl,twocolumn,showpacs,superscriptaddress,groupedaddress]{revtex4}  
\usepackage{graphicx}  
\usepackage{float}
\usepackage{dcolumn}   
\usepackage{bm}        
\usepackage{amssymb}   
\usepackage{amsmath}
\usepackage{isomath}

\begin{document}

\widetext


\title{Demonstrating an in-situ topological band transition in cylindrical granular chains}

\author{R. Chaunsali}
\affiliation{Aeronautics and Astronautics, University of Washington, Seattle, WA, USA, 98195-2400}
\author{A. Thakkar}
\affiliation{Mechanical Engineering, University of Washington, Seattle, WA, USA, 98195-2400}
\author{E. Kim}
\affiliation{Aeronautics and Astronautics, University of Washington, Seattle, WA, USA, 98195-2400}
\affiliation{Division of Mechanical System Engineering, Automotive Hi-Technology Research Center, Chonbuk National University, 567 Baekje-daero, Deokjin-gu, Jeonju-si, Jeollabuk-do, Republic of Korea, 54896}
\author{P. G. Kevrekidis}
\affiliation{Department of Mathematics and Statistics, University of Massachusetts, Amherst, MA, USA, 01003-4515}
\author{J. Yang}
\affiliation{Aeronautics and Astronautics, University of Washington, Seattle, WA, USA, 98195-2400}



\date{\today}

\begin{abstract}
We numerically investigate and experimentally demonstrate an \textit{in-situ} topological band transition in a highly tunable mechanical system made of cylindrical granular particles. This system allows us to tune its inter-particle stiffness in a controllable way, simply by changing the contact angles between the cylinders. The spatial variation of particles' stiffness results in an \textit{in-situ} transition of the system's topology. This manifests as the emergence of a boundary mode in the finite system, which we observe experimentally via laser Doppler vibrometry. When two topologically different systems are placed adjacently, we analytically predict and computationally and experimentally demonstrate the existence of a finite-frequency topologically protected mode at their interface.

\end{abstract}


\maketitle

\section{Introduction}
Topological insulators (TI)
constitute an intense area of recent interest within
condensed matter physics. TIs support directional electron transport on their surface, and this transport is immune to defects \cite{Hasan}. The existence of such a surface state has a remarkable correspondence to the non-trivial topological invariants of the bulk of TIs \cite{Qi}. Therefore, by knowing the topology of the bulk, one can predict the response on the surface/edge of the material; and this response is robust against defects on the surface as long as the topology of the bulk is preserved. This topological characterization has naturally emerged as a tool to design novel mechanical structures with unconventional elastic vibration properties on their surfaces \cite{Nash, Wang, Huber1, Mousavi,Pal}. The study of these mechanical structures not only realizes the topological phenomenon in easily controllable and observable macro-scale system, but also has potential to shape a whole new design paradigm for the structures to be used for applications, such as vibration isolation and energy harvesting \cite{Huber2}. At the same time, the paradigm of TIs has had a significant impact in other areas such as linear and nonlinear optics; see~\cite{haldane,rechtsman1,rechtsman2,ablowitz,weinstein} as examples in photonic lattices and waveguide arrays. Thus, it is a subject of broad and diverse emerging interest in phononics, photonics, acoustics, etc.

One of the most fundamental systems that contribute to our understanding of topological band theory is one-dimensional Su-Schrieffer-Heeger (SSH) dimer model \cite{SSH}. This system shows a transition across different topologies when inter- and intra-particle potentials are manipulated~\cite{SSH, Zak, Atala, Hadad}. Mathematically, this is explained by the change in a topological invariant quantity, i.e., the so-called Zak phase, of the system. Mechanical analogues of such chains have been mainly explored in context of zero-frequency topological modes~\cite{Kane}. However, mechanical systems giving rise to finite-frequency topological modes~\cite{Huber3} and potentially useful for vibration isolation purposes have been less explored, especially through experiments. This has been challenging because demonstrating a smooth topological transition within the same mechanical system requires
well controlled, \textit{in-situ} manipulation of either mass or stiffness. Changing the masses of the dimer lattice does not offer such tunability \cite{Hennion1, Hennion2}; and altering the stiffness values via external couplings, e.g., magneto-elasto \cite{Robillard}, electro-elasto \cite{Casadei}, or photo-elasto \cite{Swinteck} to name only a few, could make the system cumbersome for structural uses.  The need of precise actuation and measurement adds up to this challenge. This explains why \textit{in-situ} topological band transition for finite-frequency elastic vibrations has not been experimentally demonstrated in one-dimensional settings. A design, which can address this issue, would not only contribute to our general understanding of topological mechanical systems; it also holds promise towards
catalyzing
the implementation of stand-alone structures with tunable surface
vibration characteristics.   
 
For this purpose, we use a granular system made of cylindrical particles interacting through the Hertz contact law \cite{Johnson}. This system is highly tunable in the sense that the inter-particle stiffness can be changed simply by altering the contact angles between cylinders \cite{Khatri}. Such a versatile structure has been recently exploited for manipulating stress waves in linear \cite{Li}, linear time-dependent \cite{Chaunsali}, and non-linear media \cite{Kim}. To overcome the experimental challenges with regard to controlling the contact angles and carrying out precise measurements, we devise an experimental setup of a tunable, stand-alone cylindrical particle system. This involves
3D printed enclosures intended to support and tune the granular chain and to facilitate particles' velocity measurements through a laser Doppler vibrometer (LDV). In this way, we capture the topological band transition in the system by \textit{in-situ} control of the 1D chain's topological characteristics.
Using this tunable setup, we validate the topological band transition by detecting the emergence of an experimentally measured boundary mode in the system. We further demonstrate the existence of a topologically protected mode at the interface of two topologically distinct granular chains. Lastly, we theoretically calculate the frequency of the protected mode using symmetries in
its shape, and show that it has an excellent agreement with numerics and experiments. 

\section{Experiment and numerical setup} 
\begin{figure}[t]
\centering
\includegraphics[width=3.2in]{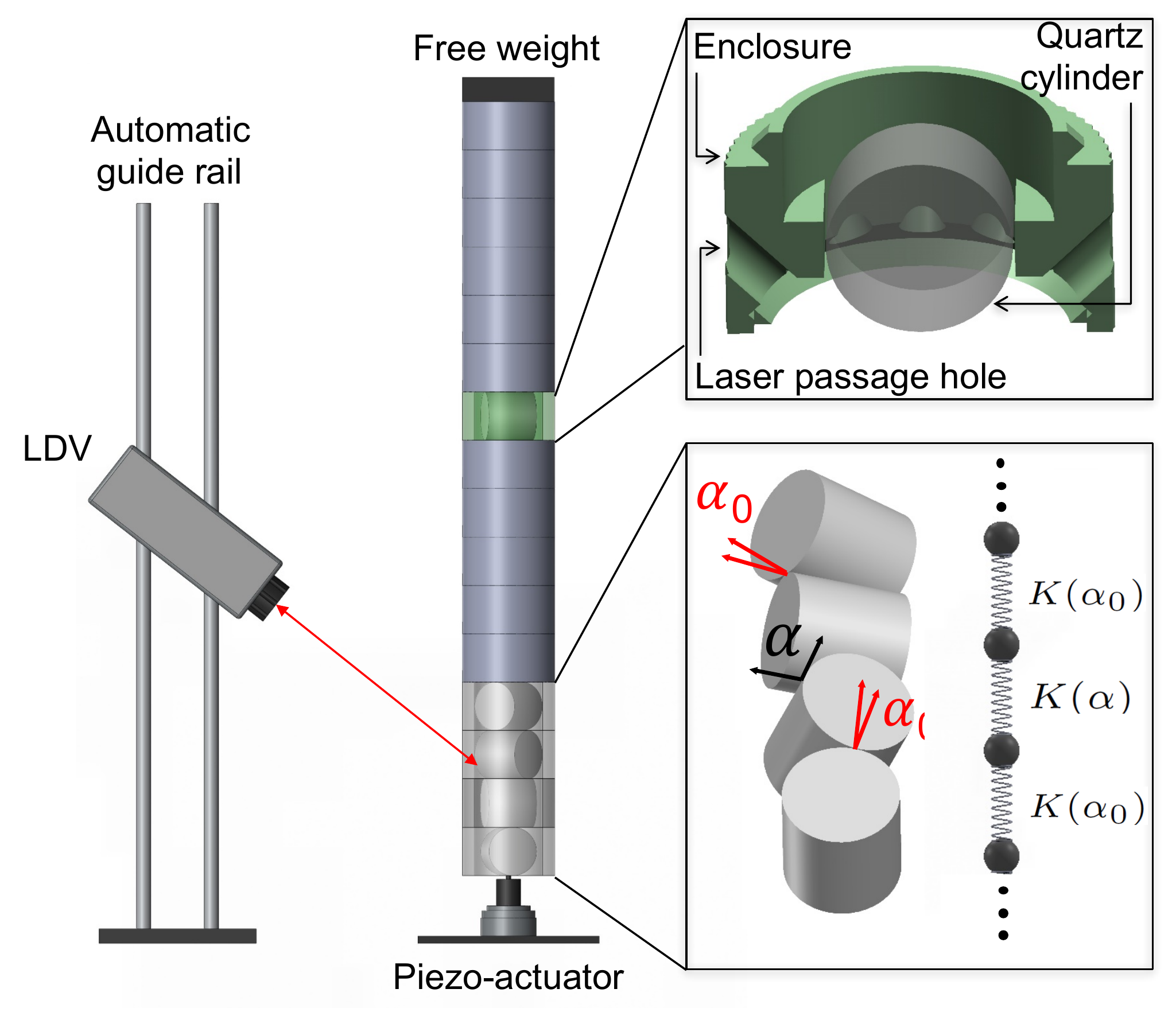}
\caption{The schematic of experimental setup used for demonstrating topological band transition in a granular chain. The top inset shows a cut section of a typical enclosure. The holes for the passage of laser beam can be seen. The bottom inset shows the contact angles in the dimer chain, and the representative spring mass system.}
\label{fig1}
\end{figure} 

The experimental setup consists of a chain of short cylinders placed inside 3D-printed enclosures stacked vertically (Fig.~\ref{fig1}). Each enclosure along with its cylinder can be independently rotated about its central axis to change the stacking angles along the chain. We maintain periodically varying two contact angles, $\alpha_0$ and $\alpha$, such that the system resembles a dimer configuration. To demonstrate the topological transition, we fix $\alpha_0$ to 20$^{\circ}$ and vary only $\alpha$ from 5$^{\circ}$ to 40$^{\circ}$. The chain is composed of 27 cylinders, and all cylinders are made of fuzed quartz (Young's modulus $Y=72$ GPa, Poisson's ratio $\mu=0.17$, and density $\rho=2187$ kg/m$^3$) with identical diameter and length of 18 mm. A piezo-actuator excites the bottom of the chain to send a frequency sweep signal from 3 kHz to 30 kHz. A free weight ($25$ N) is placed on the top of the chain to provide pre-compression to restrict the system dynamics to the linear regime. We track the velocity of each cylinder using an LDV mounted on a guide rail. Note that we have judicially designed the internal structure of the enclosures to facilitate the passage of the laser beam emanating from the
LDV at 45$^{\circ}$. 

We use a discrete element method to model the system dynamics \cite{Cundall}. In this approach, we represent the cylinders by lumped masses, and the contacts by springs following the Hertz contact law. Therefore, the force between
the $i$-th and $(i+1)$-th cylinders can be written as $F_i=\beta(\alpha_{i})[\delta_i + u_i - u_{i+1}]^{3/2}$, where $\beta(\alpha_{i})$ is the stiffness coefficient for the contact angle of $\alpha_{i}$; $\delta_i$ is the initial static compression due to the free weight;  $u_i$ and $u_{i+1}$ denote the dynamic displacements of $i$-th and $(i+1)$-th cylinders in the longitudinal direction, respectively (see the Supplemental Material for details~\cite{Suppl}). If $|u_i - u_{i+1}| \ll \delta_i $, as is the case here,
we can linearize the contact force law. Hence, contact between $i$-th and $(i+1)$-th cylinders can be assigned to a linear stiffness coefficient, i.e., $K(\alpha_i)=\frac{3}{2}\beta(\alpha_{i}) \delta_{i}^{1/2}$. This means that a dimer configuration with alternating $\alpha_0$ and $\alpha$ angles can be represented by a lumped mass model with linear stiffness coefficients, $K(\alpha_0)$ and $K(\alpha)$, varying along the chain as shown in the bottom inset of Fig.~\ref{fig1}. 

For an infinitely long dimer chain, it is straightforward to establish linear dispersion relation and calculate the edges of Bloch bands~\cite{kittel}. For a finite lattice, however, we expect to observe boundary effects. To this end, we perform
the relevant
eigenvalue analysis.
For an $N$ particle chain, we use $u=[u_1, u_2, u_3, ... , u_N]=U\exp{j\omega t}$, where $U$ and $\omega$ represent amplitude of displacement vector and angular frequency, respectively, $t$ is time, and $j$ is an imaginary unit number. Thus, by neglecting dissipation in the system, we obtain $\mathbf{\Lambda} U=\omega^2 m U$, where $m$ is the particle mass, and $\mathbf{\Lambda}$ is a $N \times N$ tridiagonal matrix consisting of stiffness coefficients, $K(\alpha_0)$ and $K(\alpha)$. This finite system also accounts for the boundary condition of the finite system. Specifically, we fix the boundaries by choosing stiffness values $K_{a}=2.78\times10^7$ N/m and $K_{w}=1.62\times10^7$ N/m at the front (actuator side) and the end of the chain, respectively, to match the experimental data (see the Supplemental Material for details~\cite{Suppl}). 
Using this finite setup, we evaluate eigenfrequencies and eigenmodes of the system in comparison with analytical and experimental data.

\section{Results and discussions}
\subsection*{Topological band transition in infinite lattice}
\begin{figure}[t]
\centering
\includegraphics[width=3.5in]{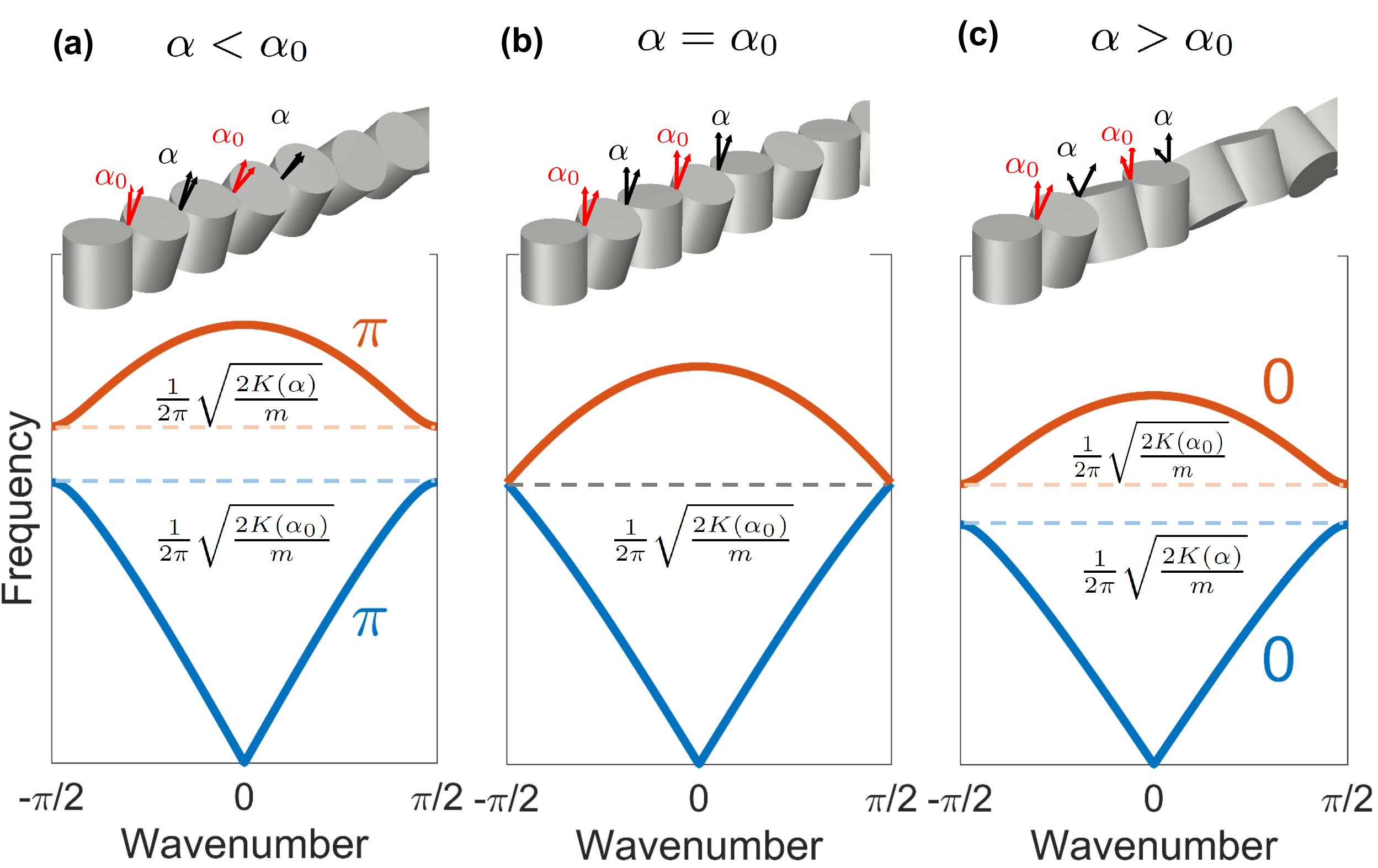}
\caption{Representative configurations of the dimer chain to show topological band transition as a function of angle $\alpha$. (a),(b), and (c) include the Bloch dispersion curves for $\alpha<\alpha_0$, $\alpha=\alpha_0$, and $\alpha>\alpha_0$, respectively. Bands are marked with the corresponding topological indices.}
\label{fig2}
\end{figure} 

Figure~\ref{fig2} includes three dimer configurations that represent a
topological band transition within our system of interest.
In the lower panel, theoretically obtained Bloch dispersion curves are plotted, in which the band-gap edges correspond to frequencies of $(1/2\pi)\sqrt{2K(\alpha_0)/m}$ and $(1/2\pi)\sqrt{2K(\alpha)/m}$. The cutoff frequency, which is the upper band edge of the optical band, equals $(1/2\pi)\sqrt{2(K(\alpha_0)+K(\alpha))/m}$. 
We notice that if $\alpha$ increases and crosses $\alpha_0$,
the frequency band-gap first closes and then opens again. This indicates a typical topological band transition in our system as a function of the angle $\alpha$. The system transitions between distinct dimer configurations that cannot be transformed to each other without closing the band-gap. The mathematical quantification of this notion can be made by calculating the topological index in one dimension -- Zak phase -- for the bands;
see the Supplemental Material~\cite{Suppl} for a relevant definition and the associated calculation.
The topological properties of a band-gap are linked to the sum of Zak phases of all the bands below the gap \cite{Xiao}. Therefore, as marked in Fig.~\ref{fig2}, the change of Zak phase from $\pi$ (for $\alpha <\alpha_0$) to $0$ (for $\alpha >\alpha_0$) for the acoustic band further supports our observation on the topological band transition in the current system.     

\subsection*{Emergence of boundary mode in finite lattice}
\begin{figure}[t]
\centering
\includegraphics[width=3.5in]{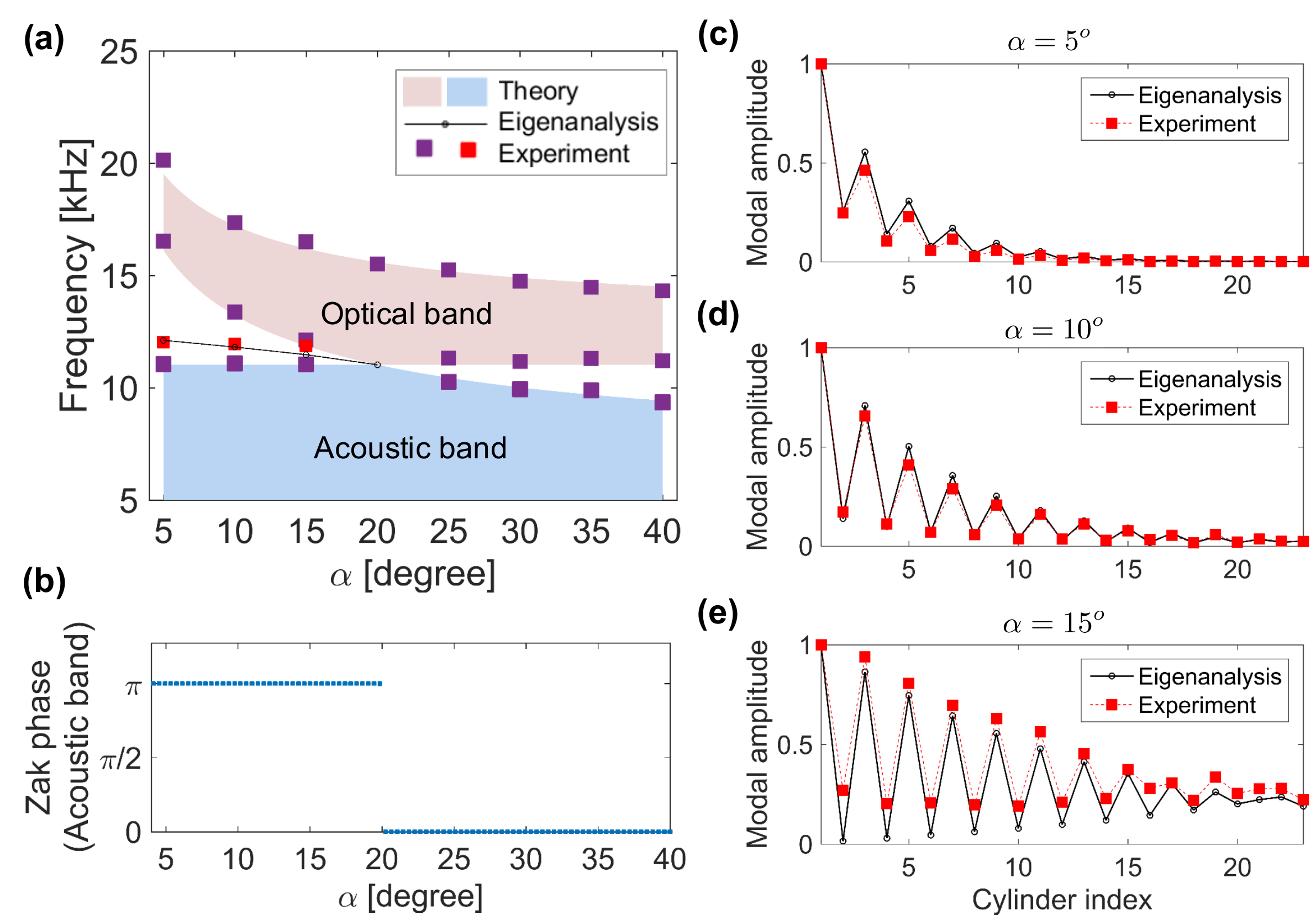}
\caption{Topological transition in the system, and emergence of boundary mode. (a) Frequency spectrum of the finite lattice. Purple markers denote the experimental data matched with the edges of bulk bands (shaded area). Red markers (experimental) show the emergence of boundary mode inside the band-gap, which follow the trend of numerical simulations (black curve). (b) Zak phase of the band-gap and its transition at $\alpha=\alpha_0$. (c-e) Numerical and experimental boundary mode profile with varying localization lengths for the configurations with $\alpha<\alpha_0$.}
\label{fig3}
\end{figure} 

The topological nature of bulk band-gaps observed for an infinite lattice manifests itself through the emergence of boundary mode(s) inside the band-gaps for a finite lattice (i.e., the principle of bulk-boundary correspondence)~\cite{Hasan}. Therefore, we expect the existence of a boundary (local/edge) mode in the current finite sized, topologically non-trivial, dimer configurations. 
To show the above phenomenon, we extract the cutoff frequencies, band-gap edges, and the local mode information from experimental data with $\alpha$ varying from $5^{\circ}$ to $40^{\circ}$ in steps of $5^{\circ}$ (see the Supplemental Material~\cite{Suppl} for more details). Figure~\ref{fig3}a summarizes the modal frequencies of the system as we vary $\alpha$. Shaded areas denote modes corresponding to the lower (acoustic) and upper (optical) branches of dispersion curves, constructed theoretically. We observe an excellent match of experimental data (purple square markers) with the edges of the theoretical bands. This successfully demonstrates the band-gap closing and opening mechanism in our tunable system. 

Due to the finite size of the chain, we experimentally observe a local mode
residing inside the band-gap (red square markers present for $\alpha < \alpha_0$). Black curve data, obtained through the numerical eigenvalue analysis, follows the same trend. These are boundary modes localized in the front of the chain, which are not witnessed when $\alpha>\alpha_0$. Emergence of these modes for all the configurations with $\alpha<\alpha_0$ suggests a change in the band topology at $\alpha=\alpha_0$. In Fig.~\ref{fig3}b, we plot the Zak phase of the acoustic band as a function of $\alpha$ to support the aforementioned argument. In Fig.~\ref{fig3}c-e, we show the extracted mode shapes (absolute values) for the boundary modes from the experiments and the eigenanalysis. These exponentially decaying profiles match closely between experiments and numerics. The localization length, however, increases as we move from $\alpha=5^{\circ}$ to $\alpha=15^{\circ}$. This can be explained by extending the intuitive arguments given in~\cite{Allen}, according to which the localization length depends on the stiffness ratio as $\xi=2/ \ln \big[ K(\alpha)/ K(\alpha_0) \big ]$. Therefore, the localization length becomes large as we increase $\alpha$ for
$\alpha<\alpha_0$, and it is natural to expect that the mode becomes
extended and is \textit{lost} inside the band in the limit
of $\alpha \rightarrow \alpha_0$. 

We note that the study of these boundary modes, i.e., the \textit{gap modes} in classical lattices, can be traced back to the pioneering work of Wallis \cite{Wallis}. He extensively explored the effect of finite boundaries and defects on the spectrum of ordered lattices in terms of generation of gap modes. However,
the novelty of the first part of this work is that we have kept the boundary conditions and the length of the system the same, and experimentally demonstrated how the change in bulk properties reflects in the emergence of a finite-frequency boundary mode as per topological band theory for one-dimensional mechanical dimers. 

\subsection*{Topological defect and protected modes}
\begin{figure}[t]
\centering
\includegraphics[width=3.55in]{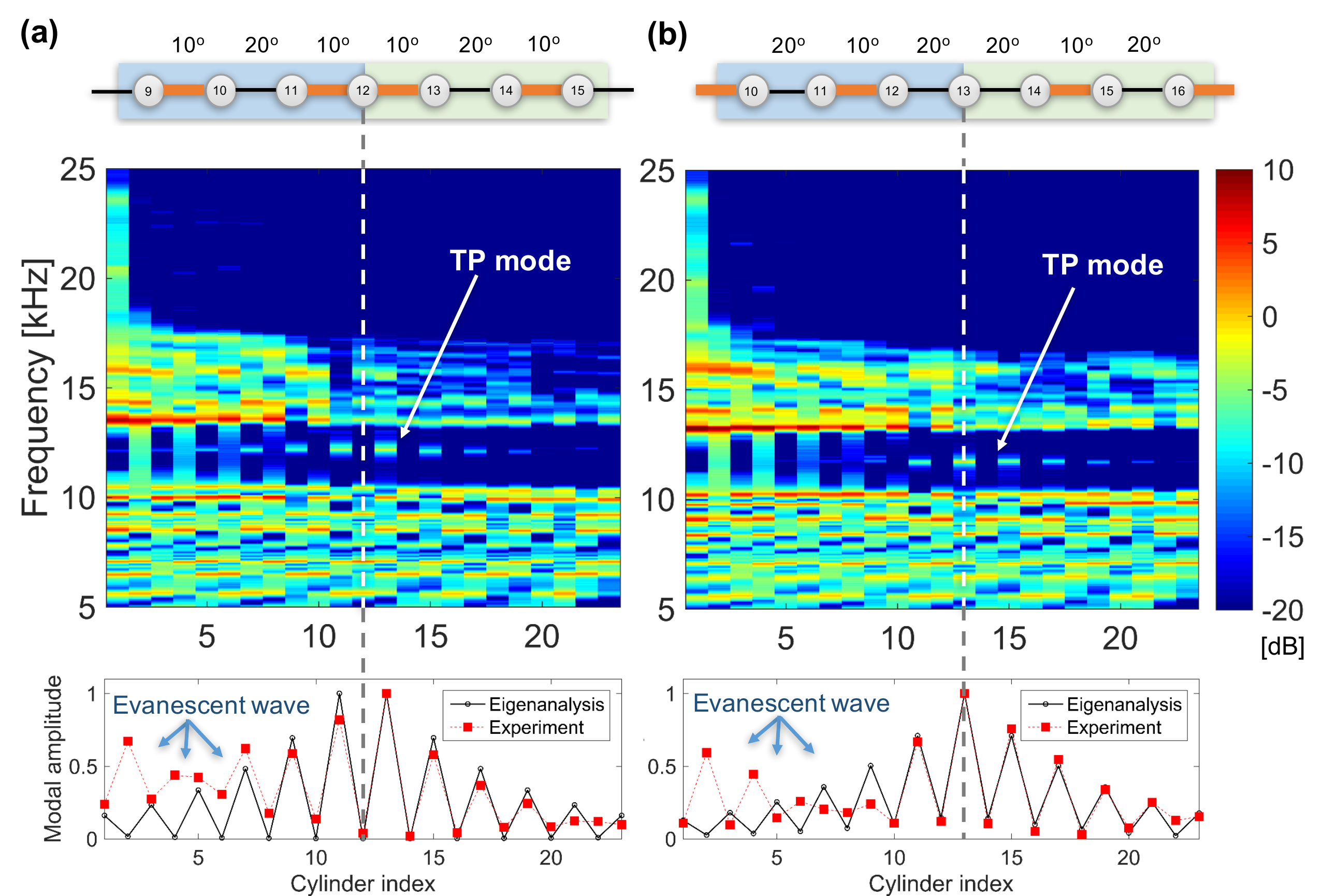}
\caption{Topologically protected (TP) vibration mode. (a) A dimer configuration with a topological defect with \textit{hard-hard} interface. Cylinder indices and contact angles are marked. Below is the spectrum plot at every space location obtained experimentally. Colors indicate power spectral density. At the bottom is the extracted TP mode shape. (b) The same with a topological defect with \textit{soft-soft} interface.}
\label{fig4}
\end{figure} 

%

We now study the ramifications of having two topologically different
dimer configurations connected to each other. For this, we assemble two dimer chains both with $\alpha_0 = 10^{\circ}$ (hard) and $\alpha = 20^{\circ}$ (soft), but connect with each other in mirror symmetry, such that we introduce a topological defect at their interface (see the top panels in Fig.~\ref{fig4}). The connecting interface can be of two types: $10^{\circ}-10^{\circ}$ or $20^{\circ}-20^{\circ}$, representing \textit{hard-hard} and \textit{soft-soft} interfaces, respectively. To understand why this defect is topological and how it is different from a trivial defect, we present the following argument. Suppose we have
the
freedom of modifying any inter-particle contact stiffness along the chain. Now, in a dimer system, a
topologically trivial defect can be introduced by going to the desired location in space and changing the stiffness locally. Similarly, the
argument can be made that if there is a trivial defect, we can go to
the defect site and perturb the stiffness locally to restore the original
defect-less dimer configuration. However, the case of a topological defect is
special  in that it cannot be straightforwardly reverted to such a defect-free
scenario through a compact perturbation. In other words, even though there exists a defect, it cannot be removed by the aforementioned \textit{local fixture} process. If one tries to move to the defect site and remove the defect by changing the local stiffness, one can see that it is not possible unless \textit{all} the stiffness values on one side of the chain are modified (i.e., a non-compact perturbation).
In essence, a topological defect can be removed only by changing the topology of one side of the configurations adjacently placed. Therefore, the
vibration mode caused by this defect is topologically protected and robust against local perturbations around the interface location. We elaborate on the effects of such local perturbation on both topologically trivial and non-trivial modes in the Supplemental Material~\cite{Suppl}.


To demonstrate the topologically non-trivial modes, we follow the same procedure as mentioned in the earlier section for the experimental study. A piezo-actuator is used for sending a frequency sweep signal from one end. It is understood that if a vibration mode caused due to a defect is localized in the middle of the chain, it is not practical to excite it using the input signal sent from the end of the chain. However, our current system is short enough that the topologically protected (TP) mode -- localized in the middle of our short chain -- can still be excited 
by coupling it to the evanescent waves inside the band-gap.  
In this way, we detect the existence of TP modes for \textit{hard-hard} and \textit{soft-soft} configurations (Fig.~\ref{fig4}). Spectrum plots are obtained experimentally by performing Fourier transformation on the temporal velocity profiles of all particles in each chain. As indicated by the arrows, we can evidently observe the existence of the TP modes in both cases. In the bottom panel, we show the extracted mode shapes (absolute values of velocities) from the experimental data, which agree well with the corresponding computed
eigenmodes. The deviation in the initial part of the chain is due to the evanescent wave in experiments, which the numerical eigenanalysis does not incorporate.

Lastly, we derive analytical expressions for the frequencies of the TP modes by utilizing the symmetries of their eigenmodes, which are evident through their spatial waveform (see the bottom panel of Fig.~\ref{fig4}). Let $K_h$ and $K_s$ denote the linear stiffness corresponding to \textit{hard} ($10^{\circ}$) and \textit{soft} ($20^{\circ}$) contacts. For the \textit{hard-hard} interface, we observe from the mode shape in Fig.~\ref{fig4} that alternate particles do not move, while the rest of the particles oscillate around their equilibrium positions. Thus, considering that their net stiffness is $K_h+K_s$, the frequency of the TP mode in this \textit{hard-hard} case would be  $f_h=\frac{1}{{2\pi}} \sqrt{\frac{K_h+K_s}{m}}$. Similarly, for the \textit{soft-soft} interface, one can show that the frequency is $f_s=\frac{1}{{2\pi}} \sqrt{\frac{2K_s(1+\gamma^{-1})}{m}}$, where $\gamma=\frac{3r-1+\sqrt{9r^2 -14r+9}}{2(r-1)}$ with $r=\frac{K_h}{K_s}$ (see the Supplemental Material for further details~\cite{Suppl}). We compare these analytical frequencies with the experimental values extracted directly from the spectrum plots in  Fig.~\ref{fig4} and the numerical ones obtained from the eigenanalysis. For the current configuration, we find the frequency of the
\textit{hard-hard} TP mode to be 12.18 kHz (Theory), 12.18 kHz (Numerics), and $12.15 \pm 0.19$ kHz (Experiment). Similarly, the frequency of the
\textit{soft-soft} TP mode is 11.80 kHz (Theory), 11.80 kHz (Numerics), and $11.70 \pm 0.08$ kHz (Experiment). Here, the standard deviations in experiments are based on the frequencies measured from all cylinder locations.

Judging from the agreement of analytical frequencies with computational and experimental results, these expressions can be used for mathematically verifying the aforementioned topological protection of interface modes. We see that for any values of $K_h$ and $K_s$ (complying  with $K_h>K_s$), the interface modes exist and the corresponding finite frequencies reside \textit{inside} the band-gap without coalescing with the bulk bands. Hence, the modes are protected gap-modes as long as we have a topological defect (\textit{hard-hard} or \textit{soft-soft}) created at the intersection of two topologically distinct dimer configurations. Again, these TP modes are robust against perturbations near the interface in contrast to trivial defect modes, and we verify such nature of the TP modes via numerics in the Supplemental Material~\cite{Suppl} (see also other related works such
as~\cite{weinstein} and references therein). 




\section{Conclusions}
In this work, we proposed a highly tunable mechanical system made of cylindrical granular particles, which can demonstrate an \textit{in-situ} topological band transition in a controllable manner.
Using non-contact laser vibrometry, we precisely captured the smooth topological transition, and showed how it leads to the emergence of a boundary mode in the system. We demonstrated the existence of topologically protected modes at the interface of two topologically distinct dimer configurations. The experimental
observations of the resulting modes are supported by the theory
(analytically identifying their frequencies) and numerics
(confirming their spatial profiles). We also confirmed that these topologically protected modes are robust under perturbations, unlike trivial defect modes observed in granular chains.

Though the current study is limited to
linear dynamics, one of the fundamental advantages of this system is that
it can be easily tuned to incorporate nonlinear effects by imposing
more substantial dynamic displacement profiles in comparison to the
static precompression. In that light, this framework can provide a promising,
experimentally accessible testbed for future studies involving
the interplay of nonlinearity and topologically protected modes. 

\section*{Acknowledgements}
We gratefully acknowledge discussions with Prof. Michael Weinstein, Columbia University and Dr. Krishanu Roychowdhury, Cornell University. J.Y. is grateful for the support from NSF (CAREER-1553202). J.K. and P.G.K. thank the support of ARO (W911NF-15-1-0604). P.G.K. also acknowledges support from the NSF-PHY-1602994, the Alexander von Humboldt Foundation
and the Stavros Niarchos Foundation via the Greek Diaspora Fellowship
Program.

\end{document}